\documentclass[twocolumn,showpacs,preprintnumbers,amsmath,amssymb]{revtex4-2}

\usepackage{graphicx}% Include figure files
\usepackage{dcolumn}% Align table columns on decimal point
\usepackage{bm}% bold math
\usepackage{mathrsfs}
\usepackage{color}
\usepackage{ulem}
\usepackage[colorlinks=true,allcolors=blue,breaklinks=true]{hyperref}
\usepackage{breakurl}

\newcommand{\nuc}[2]{{}^{#2} \mathrm{#1}}

\begin{document}

\title{Role of  Tensor Interaction as Salvation of Cluster Structure in $\nuc{Ti}{44}$ 
}% 

\author{Chikako Ishizuka}

\affiliation{
Laboratory for Zero-Carbon Energy, Institute of Innovative Research, 
Tokyo Institute of Technology, 2-12-1-N1-9, Ookayama, Meguro-ku, Tokyo, 152-8550, Japan}

\author{Hiroki Takemoto}

\affiliation{
Faculty of Pharmacy, Osaka Medical and Pharmaceutical University, 4-20-1, Nasahara, Takatsuki, Osaka, 569-1094, Japan}

\author{Yohei Chiba}
\affiliation{
Department of Physics, Osaka City University, Osaka 558-8585, Japan
}
\affiliation{
Nambu Yoichiro Institute of Theoretical and Experimental Physics (NITEP), Osaka City University, Osaka 558-8585, Japan
}
\affiliation{
Research Center for Nuclear Physics (RCNP), Osaka University, Ibaraki 567-0047, Japan
}

\author{Akira Ono}
\affiliation{
Department of Physics, Graduate School of Science,
Tohoku University,
6-3, Aramaki Aza-Aoba, Aoba-ku, Sendai 980-8578, Japan
}

\author{Naoyuki Itagaki}

\affiliation{
Yukawa Institute for Theoretical Physics, Kyoto University,
Kitashirakawa Oiwake-Cho, Kyoto 606-8502, Japan
}

\date{\today}

\begin{abstract}
\begin{description}
\item[Background]
The $\nuc{Ti}{44}$ nucleus has been known to have a $\nuc{Ca}{40}$+$\alpha$ cluster structure,
and inversion doublet structure has been observed; however, $\alpha$ cluster structure
%\sout{is} 
tends to be washed out when 
%\sout{we expand the model space and allow} 
the breaking of the $\alpha$ cluster is allowed due to the spin-orbit interaction.
Nevertheless, $\alpha$ clustering in medium-heavy nuclei is quite a hot subject recently.
\item[Purpose]
The tensor interaction has been known to play an essential role
in the strong binding of the $^4$He nucleus, which induces 
the two-particle-two-hole (2p2h) excitation.   
Since this excitation is blocked when another nucleus approaches,
it is worthwhile to show whether the tensor effect works to keep the distance between $^4$He and $\nuc{Ca}{40}$
and becomes the salvation of the clustering in $\nuc{Ti}{44}$.
\item[Methods] 
The spin-orbit effect is included in the cluster model by using
the antisymmetrized quasi cluster model (AQCM) developed by the authors. 
We have also developed an improved version of the simplified method to include the tensor contribution ($i$SMT), 
which allows us to estimate the tensor effect within the cluster model.
The competition of these two is investigated in the medium-heavy mass region for the first time.
\item[Results]
According to AQCM, the spin-orbit interaction completely breaks the $\alpha$ cluster and restores the symmetry of
$jj$-coupling shell model when the $\alpha$ cluster approaches  the $\nuc{Ca}{40}$ core.
On the other hand, $i$SMT gives a large distance between $\alpha$ and $\nuc{Ca}{40}$ due to the tensor effect.
\item[Conclusions] 
In $\nuc{Ti}{44}$, because of the strong spin-orbit and tensor contributions, two completely different 
configurations ($jj$-coupling shell model and cluster states) almost degenerate, and their mixing becomes important.
\end{description}
\end{abstract}

%\pacs{21.30.Fe, 21.60.Cs, 21.60.Gx, 27.20.+n}% PACS, the Physics and Astronomy
\maketitle

%\section{Introduction}
\par
Owing to its distinctly large binding energy,
$^4${He} becomes
a good subsystem called an $\alpha$ cluster in the nuclear systems.
Not only in the light-mass region, but also in the medium-heavy region, 
the $\alpha$ cluster structure has been investigated.
From the theoretical side,
the $\nuc{Ti}{44}$ nucleus has been predicted to have a $\nuc{Ca}{40}+\alpha$ cluster structure~\cite{PhysRevLett.57.1215}, 
and experimentally, the inversion doublet structure has been observed~\cite{PhysRevC.42.1935,Yamaya-PTPS.132.73}, which supports the existence of cluster structure.
The cluster structure of $\nuc{Ti}{44}$ is also important in nuclear astrophysics; 
$\nuc{Ti}{44}$ is one of the key elements 
for explosive nucleosynthesis in %the observation of 
core-collapse supernovae and
the $\nuc{Ca}{40}$($\alpha$,$\gamma$)$\nuc{Ti}{44}$ reaction has been considered to be the main source for 
the production~\cite{PhysRevLett.96.041102}.
\par
%\sout{However, if we assume a simple} 
In a simple description with an $\alpha$ cluster wave function,
the contribution of the non-central interactions (spin-orbit and tensor interactions) vanishes
owing to the
%\sout{antisymmetrization effect} 
spin and isospin saturation.
It is, however, widely known that one of the non-central interactions,
the spin-orbit  interaction,
is important  
in explaining the magic numbers of $28$, $50$, and $126$
of the $jj$-coupling shell model~\cite{Mayer}.
If we allow the breaking of the $\alpha$ cluster(s),
the cluster and shell structures compete with each other due to the spin-orbit interaction~\cite{PhysRevC.70.054307}.
For instance, in $\nuc{Ti}{44}$, 
when we activate the degrees of freedom of four nucleons
outside 
% of
the $\nuc{Ca}{40}$ core
based on the antisymmetrized molecular dynamics (AMD),
the $\alpha$ cluster structure 
is washed out %our 
due to the strong spin-orbit interaction,
and they perform independent particle motions~\cite{KIMURA200658,PhysRevC.93.034319}.
A similar discussion
%\sout{is performed} 
was made also for $\nuc{Ti}{48}$,
%\sout{. The}
showing that the calculated reduced width amplitude of $\alpha$ cluster structure
is the order of magnitude smaller than the one expected from the experiment~\cite{PhysRevC.103.L031305}.
Therefore, how the $\alpha$ cluster structure survives in the medium-heavy region
is quite an important subject, and many efforts have been devoted 
both from the theoretical and experimental sides~\cite{PhysRevC.81.015803,Tanaka-science}.
\par
%\sout{We} 
To quantitatively discuss the cluster--shell competition transparently, we have developed an antisymmetrized quasi cluster model
(AQCM)~\cite{PhysRevC.71.064307,PhysRevC.75.054309,PhysRevC.79.034308,PhysRevC.83.014302,PhysRevC.87.054334,ptep093D01,ptep063D01,ptepptx161,PhysRevC.94.064324,PhysRevC.97.014307,PhysRevC.98.044306,PhysRevC.101.034304,PhysRevC.102.024332,PhysRevC.103.044303},
which enables us to smoothly transform $\alpha$-cluster model wave functions to
$jj$-coupling shell model ones even within a single Slater determinant.
%\sout{The degrees of freedom of AMD are reduced in AQCM, which 
%allows us to quantitatively discuss the cluster-shell competition transparently.}
%\sout{We call the clusters that feel the effect of the spin-orbit interaction owing to this model quasi clusters, and two parameters, the distance(s) between clusters ($R$) and breaking of the cluster(s) ($\Lambda$) are needed
%to describe this competition. 
%This model is utilized to include the spin-orbit effect in the cluster model.}
The spin-orbit effect is included by breaking $\alpha$ clusters with a control parameter $\Lambda$ in this model, and such clusters that feel spin-orbit interaction are called quasi clusters. The cluster--shell competition can be described with the parameter $\Lambda$ and the distance(s) between clusters.
\par
The effect of another non-central interaction, the tensor interaction,
is included by using a different model, $i$SMT also developed by us~\cite{PhysRevC.97.014304}.
By applying $i$SMT to the four-$\alpha$ cluster structure of $^{16}$O,
the tensor interaction has been shown to increase the distances among 
$\alpha$ clusters. This is because the tensor interaction induces the 2p2h
excitation of $\alpha$ cluster(s), but this is blocked when
other $\alpha$ clusters approach,
and the attractive tensor contribution is suppressed. 
As a result, the tensor interaction works to keep the relative distances between clusters. 
This $i$SMT was reformulated as high-momentum %High-momentum 
antisymmetrized molecular dynamics~\cite{Myo-ptep-ptx143} 
AQCM-T~\cite{PhysRevC.98.054306,PTEP-Itagaki-ptz046},
and it has been possible to explain the mechanism of
the  
$\alpha$--$\alpha$ clustering in $\nuc{Be}{8}$ 
starting with a realistic nucleon-nucleon interaction.
%\sout{that it is essentially} 
An essential role is played by the tensor suppression effect at short relative distances.
\par
In this study, we discuss the competition of the spin-orbit and tensor contributions in $\nuc{Ti}{44}$.
The spin-orbit contribution is included by AQCM, which works to wash out the cluster structure.
The tensor contribution is included by $i$SMT, which enhances the relative distance between clusters.
The tensor interaction is found to be the salvation of the cluster structure.
\par
Both in AQCM and $i$SMT, the single particle wave function has a Gaussian shape \cite{Brink};
\begin{equation}
	\phi_{i} = \left( \frac{2\nu}{\pi} \right)^{\frac{3}{4}}
	\exp \left[- \nu \left(\bm{r}_{i} - \bm{\zeta}_i \right)^{2} \right] \eta_{i},
\label{Brink-wf}
\end{equation}
where $\eta_{i}$ represents the spin-isospin part of the wave function, 
and $\bm{\zeta}_i$ is a parameter representing the center of a Gaussian 
wave function for the $i$-th particle.
In the traditional cluster model (Brink model), the %\sout{$\bm{\zeta}$ value is} 
values $\bm{\zeta}_i=\bm{R}$ are common
for four nucleons with different $\eta_i=\chi^{\tau,\sigma}$, which is the definition of the $\alpha$ cluster.
We give different %\sout{$\bm{\zeta}$} 
$\bm{R}$ values for other cluster(s) if exists, 
and the $\nuc{Ca}{40}$ core %\sout{s}} 
is constructed with ten
$\alpha$ clusters with small (0.2~fm) relative distances. After the antisymmetrization, %\sout{they become}
it becomes identical to the closed-shell configuration %\sout{s} 
of the shell model.
The size parameter $\nu$ is chosen to be 
0.13~fm$^{-2}$, which gives the optimal energy ($-339.51$~MeV, with an interaction explained later) and
a reasonable radius (root mean square (rms) matter radius of 3.38~fm) for $\nuc{Ca}{40}$.
We add one $\alpha$ cluster outside the $\nuc{Ca}{40}$ core %\sout{s}
in $\nuc{Ti}{44}$.
\par
For the inclusion of the spin-orbit effect, we adopt AQCM.
The four nucleons in the $\alpha$ cluster outside of the $\nuc{Ca}{40}$ core 
is transformed into
$jj$-coupling shell model based on the AQCM,
by which the contribution of the spin-orbit interaction
due to the breaking of $\alpha$ clusters is included.
Here
the $\bm{\zeta}_{i}$ values are changed to complex numbers.
When the original value of the Gaussian center parameter $\bm{\zeta}_{i}$
is $\bm{R}$,
which is 
real and
related to the spatial position of this nucleon, 
it is transformed 
by adding the imaginary part as
\begin{equation}
  \bm{\zeta}_{i} = \bm{R} + i \Lambda \bm{e}^{\text{spin}}_{i} \times \bm{R}, 
  \label{AQCM}
\end{equation}
where $\bm{e}^{\text{spin}}_{i}$ is a unit vector for the intrinsic-spin orientation of this
nucleon. 
The control parameter $\Lambda$ is associated with the breaking of the cluster,
and two nucleons with opposite spin orientation have $\bm{\zeta}_{i}$ values
that are complex conjugate to each other.
This situation corresponds to the time-reversal motion of two nucleons.
After this transformation, the $\alpha$ cluster is called a quasi cluster.
The four nucleons are transformed to the  
 $ \left( f_{7/2} \right)^4$ configuration
of the $jj$-coupling shell model at the limit of
$|\bm{R}| \to 0$ and $\Lambda = 1$, when the $\nuc{Ca}{40}$ core is located at the origin.
\par
For the inclusion of
the rank-two non-central interaction, the tensor interaction,
we adopt $i$SMT~\cite{PhysRevC.97.014304}.
The tensor interaction
has a feature
that the second-order effect is much stronger than the first-order one;
%\sout{the effect is taken into account by incorporating}
therefore 2p2h configurations are important.
%\sout{In the original version, SMT, the real part was shifted to describe them;
%however, the effect has been quite insufficient~\cite{PhysRevC.73.034310}.
%Therefore, in the improved version of SMT, $i$SMT,
%the imaginary part of Gaussian center parameters is %shifted instead of the real part.
%This is much suited for describing the high-momentum components
%since the imaginary part of Gaussian center parameter directly corresponds to the momentum of the nucleon.}
%\par
%\red{[don't break the paragraph here] \sout{In} 
To incorporate such configurations in $i$SMT, the Gaussian center parameters for the four nucleons in $\nuc{He}{4}$
(in the example that $\nuc{He}{4}$ is located at %\sout{the origin} 
$\bm{R}$)
are %\sout{introduced} 
shifted by adding the imaginary parts in the following way;
\begin{eqnarray}
{\bm \zeta}_{p \uparrow} &=& \bm{R} + id {\bm e_z}, \nonumber \\
{\bm \zeta}_{n \uparrow} &=& %\red{\sout{0} 
\bm{R}, \nonumber \\
{\bm \zeta}_{p \downarrow} &=& \nonumber %\sout{0} 
\bm{R}, \\
{\bm \zeta}_{n \downarrow} &=& \bm{R} - id {\bm e_z}.
\label{$i$SMT-1}
\end{eqnarray}
Here, the Gaussian center parameters of proton spin-up (${\bm \zeta}_{p \uparrow} $) and neutron spin-down (${\bm \zeta}_{n \downarrow}$) are shifted,
%\sout{and $d$ values are} 
with the parameter $d$ taken to be 1.5, 3.0, $\cdot \cdot \cdot$ 15.0 fm (10 Slater determinants
in addition to $d = 0$~fm).
The spin orientation is quantized along the $z$-axis, and ${\bm e_z}$ is a unit vector for this axis.
We also prepare the basis states, where neutron spin-up (${\bm \zeta}_{n \uparrow}$) is shifted in stead of
neutron spin-down;
\begin{eqnarray}
{\bm \zeta}_{p \uparrow} &=& \bm{R} + id {\bm e_z}, \nonumber \\
{\bm \zeta}_{n \uparrow} &=& \bm{R} - id {\bm e_z}, \nonumber \\
{\bm \zeta}_{p \downarrow} &=& %\sout{0}
\bm{R}, \nonumber \\
{\bm \zeta}_{n \downarrow} &=& %\sout{0}
\bm{R}.
\label{$i$SMT-2}
\end{eqnarray}
Eventually, we superpose these 21 Slater determinants in total 
based on 
the generator coordinate method (GCM),
and coefficients for the linear combination are determined 
by diagonalizing the norm and Hamiltonian matrices.
\par
In the original version, SMT, the real part was shifted to describe them;
however, the effect has been quite insufficient~\cite{PhysRevC.73.034310}.
%Therefore, 
On the other hand, in the improved version of %SMT, $i$SMT,
SMT ($i$SMT) explained above,
the imaginary part of the Gaussian center parameters is shifted instead of the real part.
This is much suited for describing the high-momentum components
since the imaginary part of Gaussian center parameter directly corresponds to the 
nucleon momentum.
%momentum of the nucleon.
\par
The Hamiltonian ($\hat{H}$) consists of kinetic energy ($\hat{T}$) and 
potential energy ($\hat{V}$) terms,
\begin{equation}
\hat{H} = \hat{T} +\hat{V},
\end{equation}
and the kinetic energy term is described as one-body operator,
\begin{equation}
\hat{T} = \sum_i \hat{t_i} - T_{\text{cm}},
\end{equation}
and the center of mass kinetic energy ($T_{\text{cm}}$),
which is constant,
is subtracted.
The potential energy has
central ($\hat{V}_{\text{central}}$), spin-orbit ($\hat{V}_{\text{spin-orbit}}$), tensor ($\hat{V}_{\text{tensor}}$), 
and the Coulomb parts.
\par
For the central part of the nucleon-nucleon interaction 
($\hat{V}_{\text{central}}$), 
in many conventional cluster studies, the Volkov interaction~\cite{VOLKOV196533} has been often used
%\sout{for} 
as the effective nucleon-nucleon interaction.
Although this interaction is capable of describing various properties of light nuclei,
it consists of only two ranges and is not designed to well describe the 
medium-heavy region. In the present case, if we adjust the Majorana exchange
parameter %\sout{of} 
to %\sout{adjust} 
reproduce the binding energy of $\nuc{Ca}{40}$,
it gives for $\nuc{Ti}{44}$ a too deep binding between $\nuc{He}{4}$ and $\nuc{Ca}{40}$ with a very small distance between 
%\sout{$\nuc{Ca}{40}$ and $\nuc{He}{4}$} 
them, 
%\sout{for $\nuc{Ti}{44}$}
%\sout{and} 
which is not consistent with a
well-developed cluster structure,
%\sout{ does not emerge}}, 
even before taking into account
%\sout{of} 
the breaking of the $\alpha$ cluster. 
%\sout{Of course, we can 
On the other hand, if we change the Majorana parameter to
adjust the binding energy of $\nuc{Ti}{44}$ from the $\nuc{Ca}{40}+\nuc{He}{4}$ threshold energy,
%\sout{.In this case}, 
the $\nuc{Ca}{40}+\nuc{He}{4}$ cluster structure does appear
(if we do not include the breaking %effect 
of the $\alpha$ cluster),
but then the internal energy of $\nuc{Ca}{40}$ becomes very underbinding. 
The internal binding energy of $\nuc{Ca}{40}$ is indeed
not explicitly needed in the actual calculation, but the obtained results are unreliable if this value
is quite different from the experimental value.
Therefore, we need to adopt an alternative nucleon-nucleon interaction suitable for %\red{\sout{the}} 
describing the medium-heavy nuclei.
Here we introduce the one developed by 
one of the authors (H.T.),
which has three ranges;
\begin{equation}
\hat{V}_{\text{central}} = \frac{1}{2} \sum_{i \neq j}
\sum_{\alpha=1}^3 V_\alpha \exp[- (\bm r_i - \bm r_j )^2 / \mu_\alpha^2]
 (W_\alpha + M_\alpha P^r)_{ij}.
\label{2body}
\end{equation} 
Here, $P^r = -P^\sigma P^\tau$ represents the exchange of spatial part of the wave functions
of interacting two nucleons, and $W_\alpha = 1 - M_\alpha$, where $M_\alpha$ is Majorana exchange parameter.
The parameters are listed in Table~I. This interaction is designed to
reproduce the properties %\red{\sout{of} 
in a wide mass-number range.
For instance, for the symmetric nuclear matter, the interaction gives the saturation point at the binding energy %\red{\sout{par nucleon and $k_F$ value of}} 
$E/A=-15.6$~MeV and the Fermi momentum $k_F=1.29$~fm$^{-1}$. %\red{\sout{, respectively}}.
Also, it gives the incompressibility of $K = 239$~MeV. 
\begin{table}
\caption{
The parameter set for 
the central part of the nucleon-nucleon interaction 
($\hat{V}_{\text{central}}$)
adopted in the calculation (Eq.~\ref{2body}) developed by one of the authors (H.T.).}
\begin{ruledtabular}
  \begin{tabular}{cccc} 
 $\alpha$ & $V_\alpha$ (MeV)  &  $\mu_\alpha$ (fm) &  $M_\alpha$ \\ 
\hline
   1  & 611.88      & 0.81 & $ -0.06979$ \\
   2  &$-287.21$ & 1.62 & 0.55326 \\
   3  &   57.972    & 2.43 & 0.68020 \\
  \end{tabular}
\end{ruledtabular}
\label{takemoto-int}
\end{table}
\par
For the spin-orbit part ($\hat{V}_{\text{spin-orbit}}$),
Gaussian three-range soft-core (G3RS) interaction~\cite{PTP.39.91}, which is a realistic
interaction originally determined to reproduce the nucleon-nucleon scattering phase shift, 
is adopted;
\begin{equation}
\hat{V}_{\text{spin-orbit}}= \frac{1}{2} \sum_{i \ne j} V^{ls}_{ij},
\end{equation}
\begin{equation}
V^{ls}_{ij}= V_{ls}( e^{-d_{1} ({\bm r_i} - {\bm r_j})^{2}}
                    -e^{-d_{2} ({\bm r_i} - {\bm r_j})^{2}}) 
                     P(^{3}O){\bm L}\cdot{\bm S}.
\label{Vls}
\end{equation}
For the strength,
$V_{ls} = 1600\text{--}2000$ MeV has been suggested to reproduce the
scattering phase shift of $\nuc{He}{4}$ and $n$~\cite{PTP.57.866}.
We adopt $V_{ls} = 1800$ MeV, which reproduces various properties of $\nuc{C}{12}$~\cite{PhysRevC.94.064324,PhysRevC.103.044303}.
There has been discussion that the tensor contribution is included in this scattering phase shift~\cite{Arima-PTP.23.115,Myo-PTP.113.763}, and thus this strength can be considered as the maximum value.
\par
For the tensor part ($V_{\text{tensor}}$), we use Furutani interaction~\cite{PTP.60.307-Furutani}.
This interaction nicely reproduces the tail region of the one-pion-exchange potential.
\par
\begin{figure}[t]
	\centering
	\includegraphics[width=6.0cm]{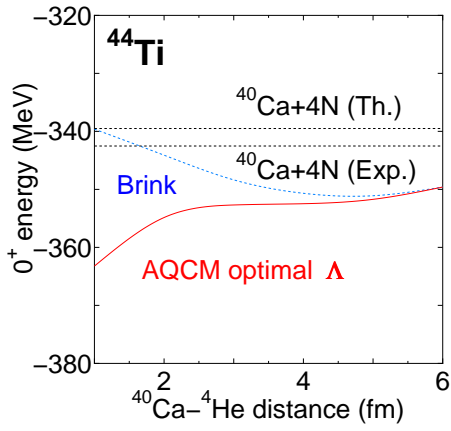} 
	\caption{
Energy curves for the $0^+$ state of $\nuc{Ti}{44}$ as a function of the distance
between $\nuc{He}{4}$ and $\nuc{Ca}{40}$.
The dotted curve is for the $\alpha$+$\nuc{Ca}{40}$ cluster model (Brink model) and
the solid curve is the result of AQCM, where the optimal $\Lambda$ value
is adopted for each distance.
}
\label{ti44-aqcm}
\end{figure}

The energy curves for the $0^+$ state of $\nuc{Ti}{44}$ are shown in 
Fig.~\ref{ti44-aqcm} as a function of the distance
between quasi cluster ($\nuc{He}{4}$) and $\nuc{Ca}{40}$.
The dotted curve is for the $\alpha$+$\nuc{Ca}{40}$ cluster model (Brink model, $\Lambda = 0$) and
the solid curve is the result of AQCM, where the optimal $\Lambda$ value
is adopted for each distance.
The horizontal dotted line at $-339.51$~MeV 
noted as $\nuc{Ca}{40}+4N$ (Th.)
shows the theoretical $\nuc{Ca}{40}+4N$ threshold energy.
Experimentally, the $\nuc{Ca}{40}$ nucleus has the energy of $-342.52$~MeV noted as $\nuc{Ca}{40}+4N$ (Exp.)
and $\nuc{Ti}{44}$ is bound from this threshold by 
32.95~MeV. There is no spin-orbit contribution for the dotted curve (Brink model), which has the %\sout{lowest}}
energy minimum around the relative distance of $\sim$4~fm,
and the binding energy there is much less than %compared with 
the experimental value (32.95~MeV below the threshold).
On the other hand, in the solid curve, AQCM, the $\alpha$ cluster can be broken by introducing the $\Lambda$ parameter, which is 
variationally determined, and the spin-orbit contribution is properly included. This effect is especially large
%\red{\sout{in} 
at the short distances between $\nuc{Ca}{40}$ and $\nuc{He}{4}$. 
The binding becomes deeper than the dotted curve (Brink model)
with decreasing the relative distance, where the binding energy approaches the experimental value.
This means that the $\alpha$ cluster structure is significantly destroyed by the spin-orbit interaction when we allow the breaking of 
the $\alpha$ cluster, and thus the result indicates that the
$jj$-coupling shell model state %\sout{is} 
may be realized in the ground state of $\nuc{Ti}{44}$.
\par
\begin{figure}[t]
	\centering
	\includegraphics[width=6.0cm]{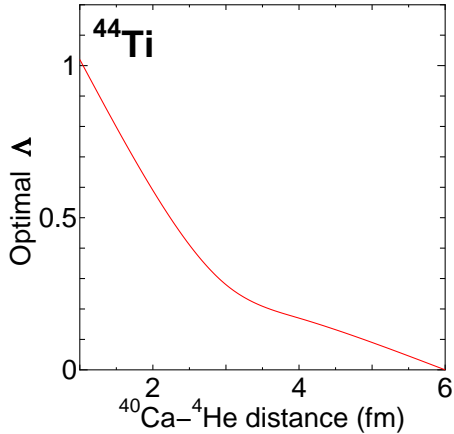} 
	\caption{
Optimal $\Lambda$ value of AQCM 
introduced to the $\nuc{He}{4}$ cluster outside of the $\nuc{Ca}{40}$ core
for the $0^+$ state of $\nuc{Ti}{44}$ 
as a function of the distance
between quasi cluster ($\nuc{He}{4}$) and $\nuc{Ca}{40}$.
     }
\label{optimal-L}
\end{figure}

This possibility is confirmed by the optimal $\Lambda$ value of AQCM, which is introduced %\red{\sout{to} 
for the $\nuc{He}{4}$ cluster outside of the $\nuc{Ca}{40}$ core.
The adopted $\Lambda$ value for the $0^+$ state of $\nuc{Ti}{44}$
is shown in Fig.~\ref{optimal-L}
as a function of the distance
between $\nuc{He}{4}$ and $\nuc{Ca}{40}$.
The value becomes almost unity at small relative distances,
and $\alpha$ cluster structure is found to be completely washed out, and the four nucleons
are changed into independent particles with the $(f_{7/2})^4$ configuration
of the $jj$-coupling shell model.
On the other hand, when the relative distance increases,
the $\Lambda$ value decreases and
the difference between the $\alpha$ cluster model and AQCM becomes small.
\par
\begin{figure}[t]
	\centering
	\includegraphics[width=6.0cm]{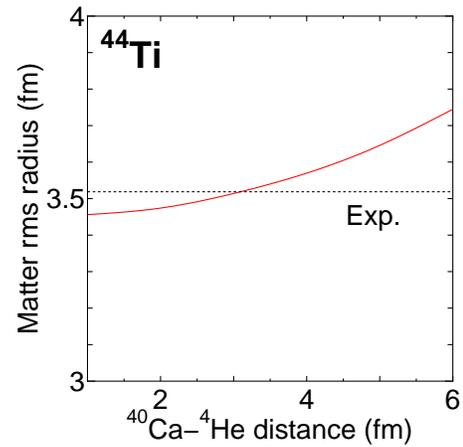} 
	\caption{
Calculated rms matter radius  
for the $0^+$ state of $\nuc{Ti}{44}$ 
as a function of the distance
between $\nuc{He}{4}$ and $\nuc{Ca}{40}$.  The optimal $\Lambda$ value of AQCM is adopted for each distance.
     }
\label{ti44-mrms}
\end{figure}

The calculated rms matter radius  
for the $0^+$ state of $\nuc{Ti}{44}$ 
is shown in Fig.~\ref{ti44-mrms}
as a function of the relative distance
between $\nuc{He}{4}$ and $\nuc{Ca}{40}$.
The optimal $\Lambda$ value of AQCM is adopted for each distance.
Experimentally, the rms charge radii of $\nuc{Ca}{40}$ and $\nuc{Ti}{44}$ are 3.4776(19)~fm and 3.6115(51)~fm, respectively.
The difference is 0.134~fm. In our model, the rms matter radius of $\nuc{Ca}{40}$ is 3.38~fm,
which is quite consistent with the experimental charge radius. 
%reproduces the experimental charge radius almost exactly.
Here we face one problem; to explain this increase of the rms radius from $\nuc{Ca}{40}$ to $\nuc{Ti}{44}$
(experimentally 0.134~fm),
the distance between $\nuc{He}{4}$ and $\nuc{Ca}{40}$ must have a certain value.
According to AQCM, the lowest energy of $\nuc{Ti}{44}$ is obtained at the relative distance of  
1~fm and $\alpha$ cluster structure vanishes, but this relative distance of 1~fm 
gives a small rms radius. The figure tells us that we need $\sim$ 3~fm as the distance between the two clusters to explain the observed increase of the rms radius
from $\nuc{Ca}{40}$ to $\nuc{Ti}{44}$. 
This means that the cluster structure must survive to some extent in the ground state of $\nuc{Ti}{44}$. 
The experimental matter radius of $\nuc{Ti}{44}$ can be deduced 
as 3.519~fm from the measured charge radius, which is shown by the dotted line, and
it crosses with the calculated result (solid line) around
the intercluster distance of 3~fm.

\par
\begin{figure}[t]
	\centering
	\includegraphics[width=6.0cm]{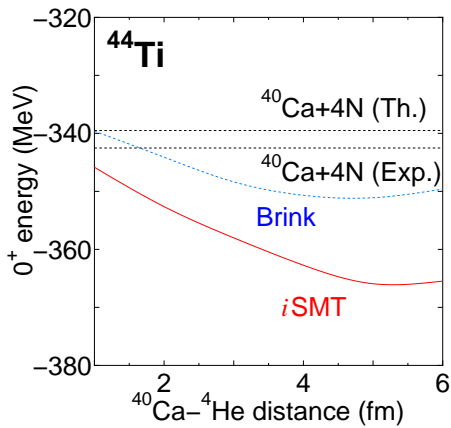} 
	\caption{
Energy of the $0^+$ state of $\nuc{Ti}{44}$ as a function of the distance
between $\nuc{He}{4}$ and $\nuc{Ca}{40}$.
The dotted curve is for the $\alpha$+$\nuc{Ca}{40}$ cluster model (Brink model) and
the solid curve is the result of $i$SMT.
     }
\label{ti44-ismt}
\end{figure}

Therefore, as the next step, we investigate the possibility of the tensor effect 
as the salvation of the cluster structure.
We introduce $i$SMT and include the tensor effect. 
The energy of the $0^+$ state of $\nuc{Ti}{44}$ 
is shown in Fig.~\ref{ti44-ismt}
as a function of the distance
between $\nuc{He}{4}$ and $\nuc{Ca}{40}$.
Here again, the horizontal dotted lines 
noted as $\nuc{Ca}{40}+4N$ (Th.)
and $\nuc{Ca}{40}+4N$ (Exp.) represent 
the theoretical and experimental $\nuc{Ca}{40}+4N$ threshold energies,
respectively.
The solid curve is the result of $i$SMT,
and again,
the dotted curve is for the $\alpha$+$\nuc{Ca}{40}$ cluster model (Brink model).
The tensor effect becomes almost zero at the limit of a small distance 
between the two clusters (tensor suppression effect), but $\alpha$ cluster model and $i$SMT give
quite different results at large relative distances. 
This means that the tensor suppression effect disappears at large relative distances.
The %\sout{lowest} 
minimum energy of $i$SMT
is given around the relative distance of 5~fm.
It can be confirmed that the tensor interaction induces the clustering of the system.
Here, the %\sout{lowest} 
minimum energy %\sout{of} 
in $i$SMT is almost the same as in AQCM, which was discussed previously.
Therefore, 
in the ground state of $\nuc{Ti}{44}$, it is considered that completely different 
structures of the $jj$-coupling shell model (AQCM) and $\nuc{He}{4}+\nuc{Ca}{40}$ clustering
($i$SMT) degenerate. If we mix two configurations with equal weight, the 
experimentally observed difference of radii
between $\nuc{Ca}{40}$ and $\nuc{Ti}{44}$ could be reproduced.
\par
Finally, we couple AQCM and $i$SMT. The AQCM is represented by the basis state
with the $\nuc{Ca}{40}\text{--}\nuc{He}{4}$ distance of 1~fm with the optimal $\Lambda$, while $i$SMT is by the distance of 5~fm.
They have the $0^+$ energies of $-362.2$~MeV and $-365.5$~MeV, and the calculated rms matter radii
are 3.46~fm and 3.63~fm.
After coupling these states, two $0^+$ states are obtained
at $-366.2$~MeV and $-362.9$~MeV. The coefficients for the linear combination of 21 basis states for $i$SMT are newly determined after coupling with the AQCM basis %\red{\sout{states,} 
state; thus it is not %\red{\sout{the} 
a numerical error that the higher state before the coupling (AQCM) is not pushed up in energy after the diagonalization of the Hamiltonian matrix. The rms matter radius of the lowest state is 3.61~fm, and the increase from the $\nuc{Ca}{40}$ one of 0.23 fm is overestimated compared to
the experimental one (0.134 fm). Experimentally, the second $0^+$ state is observed at
$E_x = 1.90$~MeV. The present level spacing of 3.3~MeV for the two $0^+$ states is larger than this value. These results would suggest that cluster configuration mixes slightly stronger than reality, and thus we have to perform more
detailed investigations. For instance, the central interaction should be reconsidered after switching on the tensor interaction. In addition, the basis states of AQCM with the relative intercluster distance less than 1~fm should be included, and more states with different intercluster distances should be superposed.
Also, the double projection has to be performed during the angular momentum projection process.
These effects are expected to reproduce more precisely the ground state energy compared to the experiment.
\par
In this article, the $\alpha$ cluster structure in the $\nuc{Ti}{44}$ nucleus was investigated, in the viewpoint of the cluster--shell competition.
The $\nuc{Ti}{44}$ nucleus has been known to have a $\nuc{Ca}{40}$+$\alpha$ cluster structure
and inversion doublet structure has been observed; however, $\alpha$ cluster structure
is washed out when we allow the breaking of the cluster and switch on the spin-orbit interaction.
Here, we focused on the role of the tensor interaction as the salvation of the cluster structure,
which has been known to play an essential role
in the strong binding of the $^4$He nucleus;
the strong tensor effect favors the large distance between the clusters.
Although the traditional $\alpha$ cluster models cannot take into account the non-central interaction,
we developed AQCM and $i$SMT to include the spin-orbit and tensor contributions in the cluster model, respectively. 
Here, %\sout{the lowest energy of $i$SMT is almost the same as that of AQCM} 
the two models predicted different ground states almost at the same energy. Therefore, 
in the true ground state of $\nuc{Ti}{44}$, completely different 
structures of the $jj$-coupling shell model (AQCM) and $\nuc{He}{4}+\nuc{Ca}{40}$ clustering
($i$SMT) are considered to mix.
After coupling AQCM and $i$SMT, two $0^+$ states are obtained
at $-366.2$~MeV and $-362.9$~MeV.  %\sout{Expeirimentally, the second $0^+$ state is observed at $E_x = 1.90$~MeV}. 
The present level spacing of 3.3~MeV is slightly larger than the experimental value ($E_x=1.90$ MeV), 
%\sout{and we have to perform more} 
which requires further
detailed investigations. 
%\red{\sout{For inst\red{a}nce, the basis states with the relative distance less than 1~fm should be {\color{red} \sout{t}}included.}}
\par
In the present study, we just added the tensor interaction in the Hamiltonian;
however, in this case, the central interaction should be reconsidered.
Also, only the angular momentum projection of the total system was performed in the present study,
but the double projection of subsystem ($\nuc{He}{4}$) and total system is expected to contribute
to the lowering of the cluster states.
These effects will be examined in the forthcoming work.
\par
Nevertheless, it is intriguing to point out that the tensor interaction works as the salvation of the cluster structure that could be otherwise destroyed by the spin-orbit interaction.
%otherwise the spin-orbit interaction could destroy.
%, which can be destroyed by the spin-obit interaction. 

\begin{acknowledgements}
Numerical  calculations have been performed at
Yukawa Institute for Theoretical Physics, Kyoto University (Yukawa-21).
\end{acknowledgements}

\bibliography{biblio_ni.bib}

\end{document}